\documentclass[aps,pra,citeautoscript,preprint,citeautoscript,showkeys]{revtex4-1}
\usepackage{amsthm}

\usepackage[main=english, catalan, american]{babel}
\DeclareUnicodeCharacter{2212}{-}
\usepackage{natbib}
\usepackage{graphicx}
\usepackage{amsmath}

\usepackage[utf8]{inputenc}
\usepackage[T1]{fontenc}
\usepackage{graphicx}
\usepackage{longtable}
\usepackage{wrapfig}
\usepackage{rotating}
\usepackage[normalem]{ulem}
\usepackage{amsmath}
\usepackage{amssymb}
\usepackage{capt-of}
\usepackage{hyperref}
\date{\today}
\title{}
\hypersetup{
 pdfauthor={LAPTOP-OD042CML},
 pdftitle={},
 pdfkeywords={},
 pdfsubject={},
 pdfcreator={Emacs 29.4.50 (Org mode 9.6.15)}, 
 pdflang={English}}
\begin{document}

\title{Comment on 'Collectively enhanced quantum measurements at the Heisenberg limit'}

\author{D. Ballester}
\affiliation{Al-Farabi Kazakh National University, 050040 Almaty, Kazakhstan}

\author{Yu. V. Arkhipov}
\affiliation{Al-Farabi Kazakh National University, 050040 Almaty, Kazakhstan}

\author{I. M. Tkachenko}
\affiliation{Departament de Matem\`atica Aplicada, Universitat Polit\`ecnica de Val\`encia, 46022 Valencia, Spain} 
\affiliation{Al-Farabi Kazakh National University, 050040 Almaty, Kazakhstan}

\date{\today}

\maketitle

In the last decades, Quantum Information Processing and Quantum
Computation (QIPQC) have boomed, promising to revolutionise our
understanding of the physical world \cite{brunner-2014-bell-nonloc,georgescu-2021-how-bell,noauthor_nobel_nodate}
and create new technologies, ranging from quantum cryptography
\cite{gisin-2002-quant-crypt} and quantum algorithms \cite{Nielsen2000-wc} to
quantum sensing \cite{giovannetti-2004-quant-enhan-measur,braun-2018-quant-enhan}.

The advantage of any QIPQC-based technology relies on two crucial
resources, i.e. quantum coherence and quantum correlations. Given how
hard it is, from a practical point of view, to generate and maintain
these resources, it is paramount to understand the performance of any
quantum application proposal with limited coherence and entanglement in
a rigorous manner \cite{braun-2011-heisen-limit,braun-2018-quant-enhan}.

D. Braun and J. Martin \cite{braun-2011-heisen-limit} started by
considering a generic system made of \(N\) quantum systems \(S_i\) coupled
to a common environment \(R\), and whose Hamiltonian can be written as

\begin{equation} \label{H-BM1}
H(x)=\sum_{i=1}^N H_i+\sum_{i,\nu}S_{i,\nu}(x)\otimes R_\nu+H_R\,
\end{equation}

where \(H_i\) is the Hamiltonian of system \(S_i\), and for simplicity the
systems \(S_i\) are taken as non-interacting; \(H_R\) denotes the
Hamiltonian of \(R\). Notice the \(x\)-dependence of the Hamiltonian Eq. \eqref{H-BM1}
through the coupling operators.

If one were to devise a way to measure how \(x\) changes through some
evolution of the quantum system, then applying quantum parameter
estimation theory, its uncertainty would be bounded by
\cite{braunstein-1994-statis-distan,braun-2011-heisen-limit}
\begin{equation} \label{dx-BM1}
\delta x=\frac{\langle \Delta A^2\rangle_x^{1/2}}{\sqrt{M}|\partial \langle
A \rangle_x/\partial x|} ,
\end{equation}
\(M\) being the number of repetitions done,
and where \(\langle A  \rangle\) is the expectation value of an observable
\(A\) in state \(\rho(x)\) and \(%
\langle\Delta A^2\rangle = \langle A^2\rangle-\langle A\rangle^2\).

Then, the authors moved to the interaction picture with respect to the
non-interacting part the Hamiltonian, \(H_0\), in Eq. \eqref{H-BM1},
i.e. \(H(x)=H_0+H_I(x)\), with \cite{braun-2011-heisen-limit}

\begin{equation} \label{H-BM2}
H_I (x)=\sum_{i,\nu}S_{i,\nu}(x)\otimes R_\nu.
\end{equation}

In order to derive an
analytical expression to show that the error \(\delta x\) would scale as
\(1/N\), they applied the second order perturbation theory to compute
\(\langle A(t)\rangle\) and \(\langle\Delta A^2(t)\rangle\) and arrived to
Eq. (13) in \cite{braun-2011-heisen-limit}. This indeed seems to
outperform the theorem derived for the unitary evolution of \(N\)
independent systems when these are prepared in an initial product state,
for which only a \(1/\sqrt N\) scaling is attainable
\cite{giovannetti-2006-quant-metrol}.

Afterwards, as a particular case, the authors of \cite{braun-2011-heisen-limit} considered
a quantum system where \(N\) resonant two-level atoms are trapped
perpendicularly in a leaky cavity mode
\cite{braun-2011-heisen-limit}. Since all couplings between the atoms and the
cavity field can be considered equal, for simplicity, the interaction
can be modelled using the Tavis-Cummings model in the interaction
picture \cite{tavis-1968-exact-solut}\}:

\begin{equation} \label{H-TC}
H_I=\sum_{i=1}^N g \left(\sigma_{i}^-a^\dagger + \sigma_{i}^+ a
\right),
\end{equation}
where \(\sigma_{i}^- =|g\rangle_{i}\langle e|_{i}\), \(\sigma_{i} ^+
=|e\rangle_{i}\langle g|_{i}\) stand for lowering and raising operators
of the two-level \$i\$-th atom, with basis
\(\{|g\rangle_{i}, |e\rangle_{i} \}\), and \(a\), \(a^\dagger\) are creation
and annihilation operators of the e.m. cavity field.

In addition, one needs to account for the spatial dependence of the
coupling \(g\) along the cavity axis \cite{Gerry2004-rc},

\begin{equation} \label{g}
g=\left(\frac{ \hbar\omega}{\epsilon_0 V} \right)^{1/2} \sin(k_z z)\,
{\mathbf{d}}\cdot {\mathbf{e}} ,
\end{equation}
where \(k_z=\pi n_z/L\), \(\epsilon_0\) denotes the vacuum dielectric constant, \(V\)
the mode volume, \({\mathbf{e}}\) is an arbitrarily oriented
polarization vector, and \({\mathbf{d}}\) is the dipole operator
\cite{Gerry2004-rc}.

The atoms are initially prepared in a dark state. As the cavity length
\(L\) changes by an amount \(\delta L\), photons would leak out through the
semi-reflecting mirror allowing to measure relative changes of the
length \(\delta L/L\).

The evolution of the system is described by the master equation
\cite{Breuer2002-tu}

\begin{equation} \label{mastter}
i \hbar \dot{\rho} = \left[H_I,\rho\right]+ i \Lambda_F \rho + i
\sum_{i}^{N} \Lambda_i \rho ,
\end{equation}
where \(H_I\) is the Hamiltonian \eqref{H-TC}, and
\begin{equation} \label{LambdaF}
\Lambda_F \rho = \kappa \left( \left[ a\rho,a^\dagger \right] + \left[
a,\rho a^\dagger \right] \right) ,$$ $$\label{Lambda_at}
\Lambda_i \rho = \Gamma/2 \left(\left[\sigma_i^- \rho, \sigma_i^+ \right] + %
\left[\sigma_i^- ,\rho \sigma_i^+ \right] \right) ,
\end{equation}
where \(\Gamma\) is
the spontaneous emission rate of each atom, and \((2\kappa)^{-1}\) is
the photon lifetime.

In order to employ the master equation above, the authors of \cite{braun-2011-heisen-limit}
used the adiabatic elimination of the photon
field to obtain an approximate solution for the reduced density matrix
of the atoms \cite{bonifacio-1971-quant-statis}, which is strictly
applicable within the superrandiance limit
\(\Gamma \ll g\sqrt N \ll \kappa\).

In order to derive the photon statistics of the photon field needed, the
well-known expressions given in \cite{bonifacio-1971-quant-statis} were
applied, leading to the following formula for the uncertainty \(\delta x\)
\begin{equation} \label{dx-BM2}
\delta x=\frac{\sqrt{2}}{\sqrt{M}gt\sqrt{N(N+2)}}\simeq\frac{\sqrt{2}}{\sqrt{%
M} gtN},
\end{equation}
for \(N\gg 1\). Despite the apparent \(1/N\) scaling, given that
superradiance occurs in the overdamped regime
\(\Gamma\ll g\sqrt{N}\ll \kappa\), for large enough \(N\), \(1/(gt) \gg \sqrt{N}/(\kappa
t)\), and
\begin{equation} \label{dx-BM3}
\delta x \simeq \frac{\sqrt{2}}{\sqrt{M} gtN} \gg \frac{\sqrt{2}}
{\sqrt{M} (\kappa t) \sqrt{N} }
\end{equation}
which corresponds to the standard quantum limit
scaling. Therefore the coupling cannot be chosen arbitrarily.

Whereas, some of practical limitations, leading to a brakedown of the
Heisenberg-limited formula for large values of \(N\) have been discussed in the
literature \cite{braun-2018-quant-enhan}, including e.g. the need to
allow for the size of the cavity to grow as N grows to keep the coupling constant, our main criticism here has to do with its own
mathematical self-consistency of the asymptotic expansion.
This is a fundamental drawback in the above derivation of
\cite{braun-2011-heisen-limit}, since it really limits the applicability of
the approach in cases of decoherence
\cite{braun-2011-heisen-limit,braun-2018-quant-enhan}. Regardless of its interesting physical
insights, it is still necessary to show that any quantum metrology
technologies proposed can outperform the standard quantum limit. Especially as the complexity of the problem grows asymptotically.

\begin{acknowledgments}
We thank Prof. Dr. Braun for fruitful discussions. This research was funded by the Science Committee of
the Ministry of Science and Higher Education of the Republic of
Kazakhstan (Grant No. AP09260349). D.B. is also grateful for the partial
support from EPSRC EP/H050434/1.
\end{acknowledgments}

\bibliography{../../bibliography/references}
\end{document}